\documentclass[journal=jpcbfk,manuscript=letter]{achemso}

\usepackage[version=3]{mhchem}

\author{Alexander Stibor}
\email{stibor@pit.physik.uni-tuebingen.de}
\author{Hannah Schefzyk}
\author{J\'ozsef Fort\'agh}
\affiliation[Uni Tuebingen]
{Physikalisches Institut, Eberhard-Karls-Universität Tübingen, CQ Center for Collective Quantum Phenomena and their Applications,
Auf der Morgenstelle 14, D-72076 Tübingen, Germany, www.pit.physik.uni-tuebingen.de/fortagh}

\title[\texttt{achemso}]
{Sublimation of the Endohedral Fullerene Er$_{3}$N@C$_{80}$}

\begin{document}

\begin{abstract}
The heat of sublimation of the endohedral metallofullerene Er$_{3}$N@C$_{80}$
was measured via Knudsen effusion mass spectrometry.
The large molecule consists of a C$_{80}$ fullerene cage which is stabilized by comprising
a complex of three erbium atoms bounded to a nitrogen atom and has a mass of 1475 amu.
The mass spectrum at a temperature of 1045 K and the relative intensities of the thermal
fractions of Er$_{3}$N@C$_{80}$ are provided. We also discuss possible thermal decomposition processes for these particles.
By measuring the quantity of evaporated molecules in thermal equilibrium
through a quadrupole mass spectrometer in a temperature range between 782 K and 1128 K, a value for the sublimation enthalpy of
$\Delta H_{sub} = 237 \pm7$~kJ~mol$^{-1}$ is obtained from the second law method.
\end{abstract}

\section{Introduction}

\begin{figure}
  \includegraphics[width=6cm]{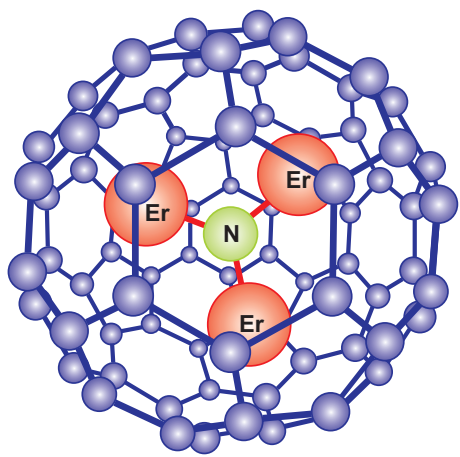}
  \caption{The chemical structure of the Er$_{3}$N@C$_{80}$ endohedral fullerene. A Er$_{3}$N nitride cluster is enclosed by a icosahedral C$_{80}$ carbon cage.}
  \label{fig1}
\end{figure}

The discovery of multimetallofullerenes within the last decade opened a new field in fullerene research with novel approaches for fundamental as well as industrial applications. These stable and large molecules can comprise a variety of atomic
dimers or even small molecules \cite{Shinohara2000, Dunsch2006}. Especially the class of trimetallic nitride endohedral fullerenes attracted attention due to their high structural stability under
ambient conditions \cite{Krause2001}, reasonable solubility in common solvents and optical properties \cite{Dantelle2009}. The enclosed particles can be rare earth ions, being good emitters in the near infrared telecommunication wavelength region\cite{Jones2006a}. They are screened by the carbon cage from chemical and physical interactions with the environment. Different nitride cluster
fullerenes can in principle be produced with selectivity up to 90\% \cite{Dunsch2004} and are commercially available in milligram quantities. Therefore these kind of endohedral fullerenes are interesting candidates for industrial and fundamental research applications. \\
\\
In most of the rare earth ions, the f-shell electrons, responsible for the optical properties, are shielded by the 6s and 5p orbitals. Although the absorption or emission lines are rather insensitive to the environment, they can be slightly shifted by local fields.  If the endohedral molecule is in an inhomogeneous solution or matrix, the lines will get broadened \cite{Jones2006a,Dantelle2009}, since each molecule has a different charge field around. Thus rare earth transition lines are expected to show extremely narrow linewidths when they are in the gas phase in ultra high vacuum. This is an interesting condition in quantum optics, allowing the laser manipulation of internal electronic states and spectroscopy with ultra-high resolution. It is therefore of importance to study the sublimation properties of endohedral fullerenes. \\
\\
In particular the trimetallic nitride template endohedral fullerene Er$_{3}$N@C$_{80}$ does fulfill the requirements for future quantum optical experiments. It was first isolated by Stevenson et al. \cite{Stevenson1999} and is commercially available in milligram quantities and high purity. Recently it was shown, that the electronic states of the encapsulated rare earth erbium ions can be directly optically excited in the infrared region \cite{Jones2006a}, since the cage exhibits transparency for wavelengths longer than 1 $\mu$m \cite{Kikuchi1994,Shinohara2000}. The molecules were frozen in a carbon disulphide (CS$_2$) matrix at cryogenic temperature, and a non-caged-mediated optical transition between the Er$^{3+}$ $^4$I$_{13/2}$ manifold and the $^4$I$_{15/2}$ manifold was excited by a tunable 1.5 $\mu$m laser \cite{Jones2006a}. This opens up the possibility to coherently manipulate the internal Er$^{3+}$ states.\\
\\
Significant effort has been made to isolate and to characterize the chemical and structural properties of this endohedral fullerenes \cite{Stevenson1999,Olmstead2000}. The chemical structure of Er$_{3}$N@C$_{80}$ is illustrated in Fig. \ref{fig1}.
The highly symmetric, icosahedral C$_{80}$ carbon cage has an isometric form which has not been isolated in an empty cage so far. It is stabilized by the adoption of the enclosed trimetallic Er$_{3}$N nitride cluster \cite{Dunsch2006}. The characterization of the chemical and structural properties indicate the Er$_3$N moiety to be planar and aligned with one cage threefold axis \cite{Jones2006b,Olmstead2000}. To understand the optical transition lines of the erbium ion and the high stability of the endohedral fullerene, it is necessary to have a qualitative picture of the charge transfer from the nitride cluster to the cage. The observed excitation lines indicate an Er$^{3+}$ charge state for each of the three erbium atoms \cite{macfarlane2001fluorescence}. Comparing this molecule to the studies concerning a similar system, namely the Sc$_{3}$N@C$_{80}$ endohedral fullerene \cite{Krause2001}, leads to the conclusion, that one electron from each erbium atom is donated to saturate the charge state of the nitrogen atom. Since the highest charge state of the erbium is +3, two further electrons from each erbium atom are transferred to the C$_{80}$ cage. It acts as a electron acceptor with a high electron affinity that can easily take up six electrons from the Er$_{3}$N complex \cite{Krause2001}. Like in Sc$_{3}$N@C$_{80}$, we assume the existence of distinct erbium-cage bonds and a strong coupling between each erbium and the central nitrogen atom.\\
\\
This article describes sublimation properties of Er$_{3}$N@C$_{80}$. We determined the enthalpy of sublimation using Knudsen effusion mass spectrometry. It is a well-established method which has been applied already for enthalpy measurements of C$_{60}$ and C$_{70}$ fullerenes \cite{Popovic1994,Pan1991,Mathews1992}. The endohedral molecules were evaporated into the vacuum in a small furnace, heated gradually up to 1128 K, and detected with a quadrupole mass spectrometer after electron beam ionization. A high-temperature mass spectrum was recorded, revealing their outstanding stability. Despite their high mass, they can be efficiently evaporated into the gas phase without significant dissociation.\\
It is an interesting question if the rearrangement of charge from the nitride cluster to the cage does change the polarizability and therefore the van der Waals binding energy between adjacent endohedral fullerenes in a heated sample powder. This could potentially lead to a deviation in the enthalpy of sublimation compared to empty fullerenes. We extracted this value from the data and compared the results to the thermodynamic properties of the nearest empty fullerenes C$_{76}$ and C$_{84}$ available in the literature \cite{Brunetti1997,Piacente1997}. The performance of our evaporation cell and the temperature measurement was determined through evaporation of C$_{60}$ molecules, yielding results with good agreement to well known literature data \cite{Popovic1994,Abrefah1992,Pan1991,Piacente1995,Mathews1992}.

\section{Experimental}

\begin{figure}
  \includegraphics[width=12cm]{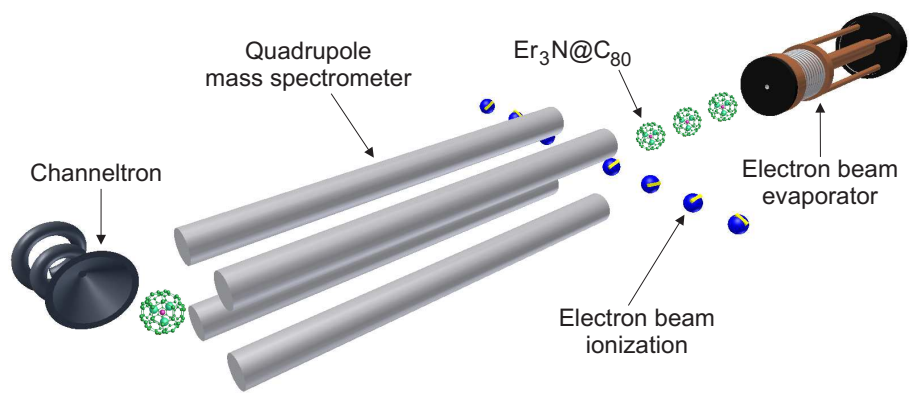}
  \caption{Experimental arrangement for measuring the sublimation enthalpy of Er$_{3}$N@C$_{80}$ endohedral fullerenes (not to scale). The molecules are evaporated into the vacuum in an electron beam evaporator and travel towards the entrance of a quadrupole mass spectrometer were they get ionized by an electron beam. Subsequently the mass selected molecules and fragments are detected by a channeltron.}
  \label{fig2}
\end{figure}

The experimental setup within an ultra-high vacuum chamber is illustrated in Fig. \ref{fig2}.
The molecules were evaporated by an electron beam evaporator with a cylindrical Knudsen effusion cell furnace \cite{Tectra}. The molybdenum crucible had an inner radius of 3~mm and a height of 3~mm. The effusion cell was covered by a molybdenum lid with a circular orifice of 1~mm in diameter. It was filled with approximately 10~mg of Er$_{3}$N@C$_{80}$ with a purity of 98\% \cite{Luna}. The remaining fraction consists of other endohedral fullerene species, mainly Er$_3$N@C$_{2x}$ where $x = 39, 41, 42, 43,$ and also Er$_2$@C$_{82}$. Residual solvents from the production process were not removed from the material. After the experiment about 1~mg fullerene powder was left in the oven.
The temperature was measured with a type-K thermocouple (NiCr-Ni) clamped close to the molecules within the wall of the molybdenum crucible. It is specified to have an accuracy of at least 1.1 K or 0.4\%.\\
\\
The evaporated endohedral fullerenes were measured with a quadrupole mass spectrometer \cite{Extrel} that was positioned axially opposite to the oven orifice, with a distance of 128 mm from the crucible. The pressure in the vacuum chamber was $6\times 10^{-8}$ mbar before the start of the measurement and raised up to $1\times 10^{-6}$ mbar during sample heating. Test experiments revealed, that the endohedral fullerenes can be efficiently evaporated up to the highest possible temperature of our furnace, which is 1182 K.\\
To measure the sublimation enthalpy, seven experimental runs were performed with the same furnace filling. During every run, the crucible was heated up in 10 to 13 temperature steps, each one lasting 10 minutes to ensure thermal equilibrium. In the first four runs, the temperature was raised between 782 K and 993 K, in the following three runs, the temperature was raised within a range of 793 to 1128 K. After each run, the sample was allowed to cool down to room temperature. At every temperature step 100 mass spectra, scanning with a resolution of 31 samples per mass and covering the range between 200 amu and 1550 amu, were taken and averaged. For the ionization of the molecules an electron beam energy of 200 eV was used.
In the analysis of the data, the background of each spectrum was determined and removed individually by averaging the signal between 1505 amu and 1550 amu, were no peaks arise. The total ion intensity was determined by integrating over the mass over charge (m/z) peak of the single charged endohedral fullerene, therefore between $m/z = 1465-1490$.\\
\\
The number of detected ions in the mass spectrometer $N_{ion}$ depends on the particle flux $\frac{dN}{dt}$ of the incoming molecular beam and on the time the molecules spend in the ionization region. It is thus indirect proportional to the velocity. Since the velocity is proportional to the square root of the particle's temperature $T$, the total ion signal is related to: $N_{ion}\propto\frac{\frac{dN}{dt}}{\sqrt{T}}$\\
The molecular flux at the entrance lens of the mass spectrometer depends on the gas flow regime in the furnace. It is determined by the molecular mean free path in the Knudsen cell and by the crucible dimensions. The important parameter is the Knudsen number, which is the ratio between the mean free path and the size of the oven aperture ($K=\Lambda/d$). For $K>8$ the flow is molecular, whereas for $K<0.1$ the flow is hydrodynamic \cite{Bird1994}. However, the functional relation between the flux $\frac{dN}{dt}$, the vapor pressure $p$ and the temperature $T$ is the same for both regimes \cite{Stefanov2004}: $\frac{dN}{dt}\propto\frac{p}{\sqrt{T}}$.
Nevertheless this relation is not valid in the transition region between the two regimes. The total time integrated ion intensity measured in the mass spectrometer can therefore be evaluated as: $N_{ion}\propto\frac{p}{T}$.\\
The vapor pressure can be expressed using the Clausius-Clapeyron equation: $p=A \ exp\left(-\frac{B}{T}\right)$.
$A$ and $B$ are experimental parameters and according to the van't Hoff equation, $B$ is connected to the sublimation enthalpy $\Delta H_{sub}$ as $B=\frac{\Delta H_{sub}}{R}$, with $R$ being the gas constant.\\
Finally, the measured data is plotted as $ln\left(N_{ion}T\right)$ versus $1/T$ and the slope of a least squares fit yields the average heat of sublimation:
\begin{equation}
ln\left(N_{ion}T\right) = \tilde A-\frac{\Delta H_{sub}}{R}\frac{1}{T}
\end{equation}
where $\tilde A$ includes $\tilde A=ln\,A-ln\,C$ with $A$ and $C$ being unknown proportionality constants.\\

\section{Results and discussion}

\begin{figure}
  \includegraphics[width=11cm]{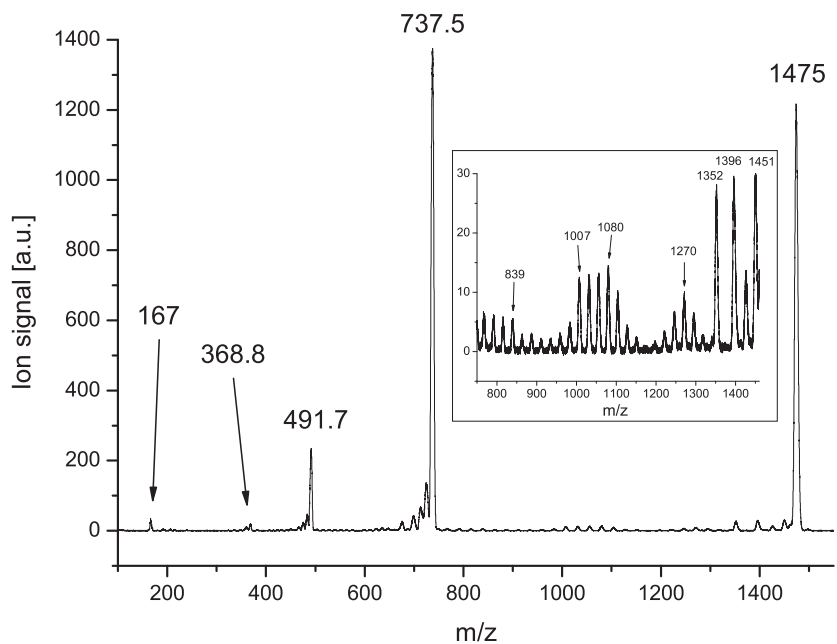}
  \caption{Average of 210 mass spectra with the four endohedral fullerene charge states Er$_{3}$N@C$_{80}^+$, Er$_{3}$N@C$_{80}^{2+}$, Er$_{3}$N@C$_{80}^{3+}$ and Er$_{3}$N@C$_{80}^{4+}$ at a temperature of 1045 K. The corresponding peaks are located at m/z values of 1475, 737.5, 491.7 and 368.8, respectively. The small peak at 167 belongs to free erbium atoms. The resolution was 18 samples per mass. The inset exhibits a magnification of the mass peaks in the m/z-region between 750 and 1460, with some indications to major peaks.}
  \label{fig3}
\end{figure}

To prove, that the Er$_{3}$N@C$_{80}$ endohedral fullerenes can be efficiently evaporated into the gas phase, an averaged mass over charge spectrum was recorded in a mass range between 100~amu and 1550~amu and at a furnace temperature of 1045~K, which is shown in Fig. \ref{fig3}. Even at this high temperature, most of the molecules do not disrupt, despite their complex structure, high mass and the high energy electron ionization. The extraordinary stability of this cluster is also demonstrated by the peak of free erbium at $m/z = 167$ in the mass spectra of Fig. \ref{fig3}, having a small relative intensity of only 21 compared to 1000 for the peak of the twofold charged endohedral fullerenes (see Table 1). Additionally it has to be noted, that with each broken cage three erbium atoms are released and that the presence of free erbium, originating from the sample production process, is unlikely \cite{LunafreeErbium}. This is an indication of the small amount of C$_{80}$ shells that got decomposed through the thermal excitation.
As expected, there is no evidence for empty C$_{80}$ cages due to the absence of a peak at 960 amu.\\
\\
Several other small peaks are observed that represent fragments of the Er$_{3}$N@C$_{80}$ molecule, mainly within mass steps of multiples of 24 amu. This is an indication for thermal emission of parts comprising multiples of carbon pairs. Their determined m/z values and their maximal intensities relative to the twofold charged Er$_{3}$N@C$_{80}^{2+}$ peak are listed in Table 1. Most of the peaks can be explained by 4 decomposition processes. But it has to be noted, that the width of the peaks allows only a resolution within $\pm$1~amu. For that reason other mechanisms may also be possible to create the observed mass spectrum. The 13 peaks within the mass range of 767 amu to 1055 amu can be assigned to the separation of Er$_2$N\,C$_{2n}$ with $n$ being an integer between 3 and 15. It may also be explained by the partition of C$_{1+2n}$ ($n=17-29$). The peaks at the m/z values of 1128, 1104 and 1080 can be created through the loss of Er\,C$_{15}$, Er\,C$_{17}$ and Er\,C$_{19}$. The 6 peaks between 1294 and 1150 amu can be due to the separation of Er\,NC$_{2n}$ ($n=0-6$), whereas the one around 1174 amu is slightly higher than the noise level and is therefore not listed in Table 1. Finally the peaks at 1451 and 1427 amu can be explained by the emission of one and two C$_2$ molecules and, for the first value, the contribution of the already mentioned Er$_3$N@C$_{78}$ impurity. Only the origin of the remaining 2 peaks around 1396 and 1352 amu is unclear. The small peak around 1318 amu can be assigned to the Er$_2$@C$_{82}$ impurity within the sample.
\begin{table}
  \begin{tabular}{cc|cc|cc}
    \hline
    m/z & rel. intensity & m/z & rel. intensity & m/z& rel. intensity  \\ \hline \hline

    1475 & 864 & 1150 & 1  & 887 & 2 \\
    1451 & 21  & 1128 & 2  & 863 & 2 \\
    1427 & 10  & 1104 & 7  & 839 & 4 \\
    1396 & 21  & 1080 & 10 & 815 & 4 \\
    1352 & 20  & 1055 & 10 & 791 & 4 \\
    1318 & 1   & 1031 & 9  & 767 & 4 \\
    1294 & 4   & 1007 & 9  & 737.5 & 1000 \\
    1270 & 7   & 983  & 3  & 492   & 166 \\
    1246 & 5   & 959  & 1  & 369   & 12 \\
    1222 & 2   & 935  & 1  & 167   & 21 \\
    1198 & 1   & 911  & 1  &       & \\
    \hline
  \end{tabular}
  \caption{Mass over charge (m/z) peaks indicating fullerene fragments left after thermal decomposition. Four possible splitting processes are described in the text, explaining most of the peaks. The majority of the parts have mass differences of 24 amu, indicating the loss of C$_2$ molecules. Only single charged peaks above m/z = 737 are listed, together with the twofold, threefold and fourfold charged Er$_{3}$N@C$_{80}$ molecular peak and the atomic erbium signal. Their maximal intensities are normalized to the twofold charged peak at m/z equal to 737.5.}
\end{table}

\begin{figure}
  \includegraphics[width=16cm]{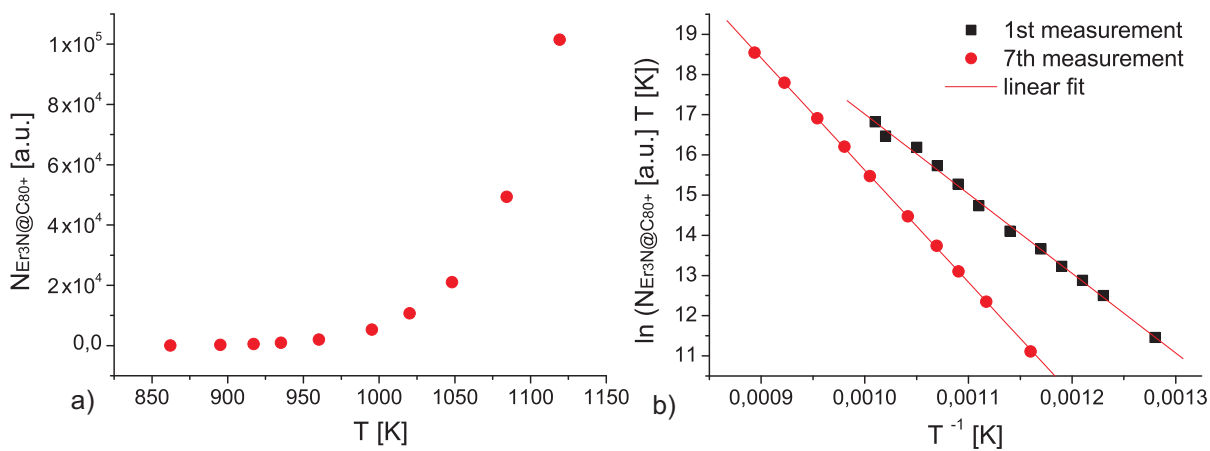}
  \caption{a) The temperature dependence of the integrated mass spectrometer signal of the single charge state of the Er$_{3}$N@C$_{80}$ endohedral fullerene in the seventh evaporation run (circles) within a temperature range between 862 K and 1119 K. b) The same data (circles) plotted in a logarithmic way to reveal the enthalpy of sublimation from the slope of a linear least squares fit (straight line) described by $ln\left(N_{Er_{3}N@C_{80}^+}\,T\right) = 44 - 27917 \, T^{-1}$. Additionally the data from the first measurement run (squares) is shown, yielding a significantly lower value for the enthalpy of sublimation probably due to a fractional decomposition of the fullerenes into amorphous carbon.}
  \label{fig4}
\end{figure}

\begin{figure}
  \includegraphics[width=8cm]{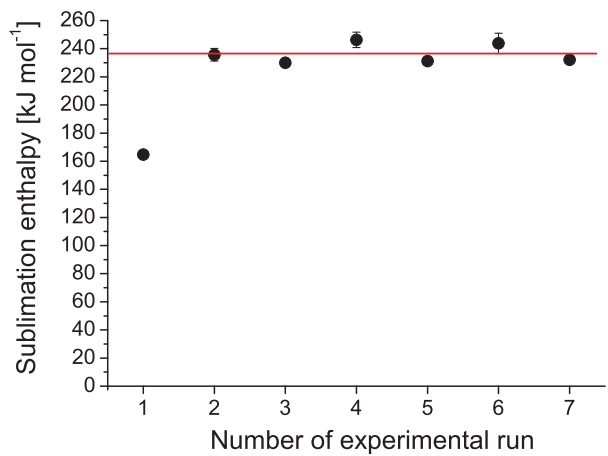}
  \caption{The determined enthalpy of sublimation for Er$_{3}$N@C$_{80}$ fullerenes in each experimental run (circles). The outcome was extracted from the slope of the logarithmically plotted evaporation data and the error bars correspond to the error in the least squares fit. Due to an assumed dependency of the enthalpy to the heat pretreatment of the sample as known for C$_{60}$ \cite{Popovic1994,Piacente1995}, the first measurement was not included to get the average value of $\Delta H_{sub, Er_{3}N@C_{80}}= 237 \pm7$ kJ mol$^{-1}$, indicated by a straight line.}
  \label{fig5}
\end{figure}

It is of interest to compare the thermal stability of Er$_{3}$N@C$_{80}$ with the one of Sc$_{3}$N@C$_{80}$, since they have very similar structure. Krause $et \,al.$~\cite{Krause2001} analyzed experimentally the structure and stability of Sc$_{3}$N@C$_{80}$ by temperature-dependent Raman and infrared spectroscopy as well as by quantum-chemical calculations. They dropcoated the 	molecules dissolved in toluene on a gold-covered silicon substrate. After sample heating and placement in ultra-high vacuum, the sample was excited by a 514~nm and 647~nm laser and the scattered Raman spectrum was analyzed by a spectrometer. The sample was heated and spectroscopic data was recorded from 80 to 700 K. After the temperature exceeded 650 K an irreversible shift in the measured frequency happened which they attribute to thermal decomposition of a part of the sample. It is remarkable, that their crystalline material on the surface showed thermal stability only up to a temperature of about 650~K, a value we did not even see significant evaporation of Er$_{3}$N@C$_{80}$, which started around 770 K.\\
\\
After this first measurement the furnace was refilled and the evaporation series, to measure the enthalpy of sublimation, was started as described above.
In Fig. \ref{fig4} a) a typical set of data (from the seventh experimental run) is shown. The integrated Er$_{3}$N@C$_{80}^+$ signal ($N_{Er_3N@C_{80}^+}$) is growing exponentially as the temperature increases from 862 K to 1119 K. The same data plotted in a logarithmic way such as $ln\left(N_{Er_3N@C_{80}^+}T\right)$ versus $1/T$ is exhibited in Fig. \ref{fig4} b) together with the data from the first run. The sublimation enthalpy obtained from the slope of a linear least squares fit was determined to be $\Delta H_{sub,Er_{3}N@C_{80}, 1st}= 165 \pm 3$ kJ mol$^{-1}$ for the first and $\Delta H_{sub, Er_{3}N@C_{80}, 7th}= 232 \pm 1$ kJ mol$^{-1}$ for the seventh run. The sublimation enthalpies extracted out of all seven evaporation curves are summarized in Fig. \ref{fig5}, indicating, that the results from the second to the sixth experimental runs reveal similar enthalpies as the seventh measurement. We performed additional series of evaporation measurements yielding always a noticeable lower enthalpy value in the first run.\\
\\
\\
In fact, the observation of a significantly lower sublimation enthalpy in the first sample heating process for other fullerenes, such as for C$_{60}$, is a well known effect in the literature \cite{Popovic1994,Piacente1995}. It is assumed, that the fullerene enthalpy depends on the heat pretreatment of the sample. The presence of small amounts of residual (organic) solvents in the original crystalline material initiates a fractional decomposition of the fullerenes into amorphous carbon. This process happens when the probe is heated for the first time and leads to the observed increased values for the enthalpy of sublimation in subsequent measurements with the same sample.\\
Following this argument, we ignored the first enthalpy value and averaged over the outcome of the other six measurements yielding a sublimation enthalpy for the endohedral Er$_{3}$N@C$_{80}$ fullerene of $\Delta H_{sub, Er_{3}N@C_{80}}= 237 \pm7$ kJ mol$^{-1}$, whereas the error indicates the standard deviation of these values. This result is compared to the literature values of the empty fullerenes C$_{60}$, C$_{70}$, C$_{76}$ and C$_{84}$ in Table 2. There are several potential reasons that contribute to the significantly increased enthalpy of sublimation for the endohedral molecule. One cause can be the notedly higher mass due to the presence of the three erbium atoms. Er$_{3}$N@C$_{80}$ has a mass of 1475~amu compared to 1008~amu of the empty C$_{84}$ fullerene. On the other hand, a modified van der Waals binding energy between the fullerenes in the sample powder can also result into a variation of the sublimation enthalpy. As mentioned in the introduction, six electrons are transferred from the erbium atoms to the cage \cite{macfarlane2001fluorescence}. This alters significantly the charge state of the endohedral fullerene cages compared to the empty ones, leading to a possible modification in the binding interaction between the molecular clusters in the sample. Furthermore, the tight binding and stabilization of the C$_{80}$ cage through the nitride cluster causes a different thermal behavior, since the rotational and vibrational modes are modified compared to the empty fullerenes. It is therefore reasonable to assume a strong variation in the Er$_{3}$N@C$_{80}$ evaporation properties. \\
\begin{table}
  \begin{tabular}{cccc}
    \hline
    Fullerene & $\Delta$T [K] & $\Delta H_{sub}$ [kJ mol$^{-1}]$ & ref. \\ \hline
    \hline
    Er$_{3}$N@C$_{80}$ & 862--1119 & 237 $\pm$7 & this work \\
    C$_{84}$ & 920--1190 & 210 $\pm$6 & \cite{Piacente1997}\\
    C$_{76}$ & 851--1069 & 194 $\pm$4 & \cite{Brunetti1997}\\
    C$_{70}$ & 783--904  & 190 $\pm$3 & \cite{Piacente1996}\\
    C$_{60}$ & 667--830  & 166 $\pm$2 & this work \\   \hline
  \end{tabular}
  \caption{Comparison between the sublimation enthalpies for the endohedral Er$_{3}$N@C$_{80}$ and the closest empty fullerenes C$_{76}$ and C$_{84}$ together with the data for C$_{60}$ and C$_{70}$.}
\end{table}
\\
To prove the good performance of our evaporation cell, we determined the sublimation enthalpy of C$_{60}$ fullerenes in the same way as for the Er$_{3}$N@C$_{80}$ molecules and compared the outcome to the literature. Our measurement series consisted of three runs with 8 temperature steps each in the range between 659~K and 830~K. All averaged mass spectrometer data in the mass range from 713~amu to 725~amu was integrated, comprising the single charged $C_{60}$ peak at 720~amu. The values for the enthalpy of sublimation were extracted to be $\Delta H_{sub, C_{60}, 1st}= 151 \pm2$ kJ mol$^{-1}$, $\Delta H_{sub, C_{60}, 2nd}= 159 \pm1$ kJ mol$^{-1}$ and $\Delta H_{sub, C_{60}, 3rd}= 166 \pm2$ kJ mol$^{-1}$ in the first, second and third run, respectively. Again it can be observed, that the first value is lower than the others.\\
For that reason we consider the sublimation enthalpy measured in the second and third run to be more representative. The data is in good agreement with other Knudsen effusion studies performed by Popovi\'c $et \,al.$ \cite{Popovic1994} ($\Delta H_{sub, C_{60}}= 158 \pm3$ kJ mol$^{-1}$), Abrefah $et \,al.$ \cite{Abrefah1992} ($\Delta H_{sub, C_{60}}= 159 \pm4$ kJ mol$^{-1}$) and Pan $et \,al.$ \cite{Pan1991} ($\Delta H_{sub, C_{60}}= 168 \pm5$ kJ mol$^{-1}$). Piacente $et \,al.$ \cite{Piacente1995} and Mathews $et \,al.$ \cite{Mathews1992} determined slightly higher values ($\Delta H_{sub, C_{60}}= 175 \pm3$ kJ mol$^{-1}$ and $\Delta H_{sub, C_{60}}= 181 \pm2$ kJ mol$^{-1}$, respectively). A list of some experimental data compared with a theoretical model is provided by Zubov $et \,al$ \cite{Zubov1997}.\\
\\
The reason for the large variations of the literature values can be mainly assigned to the pronounced effect of sample heat pretreatment to the reproducibility of the results, as discussed in detail by Popovi\'c $et \,al.$ \cite{Popovic1994} and Piacente $et \,al$ \cite{Piacente1995}. The formed amorphous carbon at high temperatures, through the disintegration of the fullerenes, has a large volume and surface area. It attracts undecomposed C$_{60}$ molecules. It is claimed, that this adsorption is the reason for the lower vapor pressure and higher heat of sublimation found in the following experimental runs after sample heating. This influence in the outcome depends on the geometry of the crucible, the maximal temperature in previous runs and on the time, the sample stayed at this maximum. All these parameters are probably chosen differently throughout the literature.\\
Another effect to be considered, is the temperature dependency of the enthalpy of sublimation, as indicated for C$_{60}$ by theoretical calculations of Zubov $et \,al$ \cite{Zubov1997}. Thus it is usual to assign the determined enthalpy to the average temperature in the heating interval. Our outcome for the Er$_{3}$N@C$_{80}$ fullerenes of $\Delta H_{sub, Er_{3}N@C_{80}}= 237 \pm7$ kJ mol$^{-1}$ corresponds to an average temperature of 955~K. For the C$_{60}$ measurement the value of $\Delta H_{sub, C_{60}}= 166 \pm2$ kJ mol$^{-1}$ corresponds to an average temperature of 745 K. Both values are evaluated by the second law method.

\section{Conclusion}

We performed an evaporation study of the endohedral metallofullerene Er$_{3}$N@C$_{80}$ via Knudsen effusion mass spectrometry. This method is advantageous compared to other approaches, like the torsion-effusion method, since we are able to selectively consider only unbroken molecules. It provides constant information about the intactness of the evaporated molecules and allows us therefore to safely go to high temperature regions.
We tested the quality of our measurement through the evaporation of C$_{60}$ fullerenes and compared the outcome to the literature. Our determined C$_{60}$ sublimation enthalpy of 166 kJ mol$^{-1}$ is in good agreement with published data, noting that the literature values show a significant spread. It is argued, that the large variations are probably due to the different heat pretreatments of the samples. This results in a notably lower measured sublimation enthalpy in the first heating run compared to subsequent runs. We observe this effect for Er$_{3}$N@C$_{80}$ as well as for C$_{60}$. \\
\\
An important outcome of our study is, that the Er$_{3}$N@C$_{80}$ fullerene is extraordinarily stable, in spite of its high mass of 1475 amu and its complex endohedral structure. It can be well evaporated up to a temperature of 1182 K and electron beam ionized with an ionization energy of 200~eV. To prove the thermal stability, we provided the relative peak heights in the mass spectrum at 1045 K.
The stability was compared to Sc$_{3}$N@C$_{80}$, a molecule with similar structure. It was heated on a surface by Krause $et \, al$~\cite{Krause2001}, showing molecular disintegration already at a temperature around 650 K. The studies support our conclusion, that Er$_{3}$N@C$_{80}$ exhibits a stability, rather unique for molecular complexes in this high mass regime. It makes this fullerene a promising candidate for matter-wave interferometry with heavy molecules \cite{Gerlich2007,Hackermueller2003}. In this field of research, the quest for proving the wave nature of molecules as heavy as possible, is inhibited by the difficulties to form and detect an intensive beam of heavy and neutral particles in the vacuum. \\
\\
Finally the enthalpy of sublimation for Er$_{3}$N@C$_{80}$ in a temperature range between 782 K and 1128 K was determined to be $\Delta H_{sub}= 237 \pm7$ kJ mol$^{-1}$. This value is significantly higher than the enthalpies of the next closest stable empty fullerenes C$_{76}$ and C$_{84}$. For that reason it can be assumed, that the considerably higher mass and the large charge transfer between the threefold charged erbium atoms and the cage does modify the evaporation characteristics. The fact that the electronic state of the enclosed erbium ions can be directly excited by a laser \cite{Jones2006a} makes Er$_{3}$N@C$_{80}$ an interesting candidate for quantum optical experiments, where the endohedral fullerenes may be sublimated and laser manipulated in the gas phase.

\begin{acknowledgement}

The authors thank Luna Innovations Inc. providing the Er$_{3}$N@C$_{80}$ fullerenes and acknowledge Andreas Günther for helpful discussions.

\end{acknowledgement}

\end{document}